\documentclass[final]{pasj01}
\usepackage{natbib}
\begin{document} 
\Received{2017/05/31}%{yyyy/mm/dd}
\Accepted{2017/08/26}%{yyyy/mm/dd}
%\Published{yyyy/mm/dd}

\title{GOLDRUSH. III. A Systematic Search of Protoclusters at $z\sim4$ Based on the $>100\,\mathrm{deg^2}$ Area}

\author{Jun \textsc{Toshikawa}\altaffilmark{1,2}, Hisakazu \textsc{Uchiyama}\altaffilmark{2,3},
    Nobunari \textsc{Kashikawa}\altaffilmark{2,3}, Masami \textsc{Ouchi}\altaffilmark{1},
    Roderik \textsc{Overzier}\altaffilmark{4}, Yoshiaki \textsc{Ono}\altaffilmark{1},
    Yuichi \textsc{Harikane}\altaffilmark{1,5}, Shogo \textsc{Ishikawa}\altaffilmark{2,3},
    Tadayuki \textsc{Kodama}\altaffilmark{2,3}, Yuichi \textsc{Matsuda}\altaffilmark{2,3},
    Yen-Ting \textsc{Lin}\altaffilmark{6}, Masafusa \textsc{Onoue}\altaffilmark{2,3},
    Masayuki \textsc{Tanaka}\altaffilmark{2}, Tohru \textsc{Nagao}\altaffilmark{7},
    Masayuki \textsc{Akiyama}\altaffilmark{8}, Yutaka \textsc{Komiyama}\altaffilmark{2,3},
    Tomotsugu \textsc{Goto}\altaffilmark{9}, Chien-Hsiu \textsc{Lee}\altaffilmark{10}}
\email{toshijun@icrr.u-tokyo.ac.jp}

\altaffiltext{1}{Institute for Cosmic Ray Research, The University of Tokyo, Kashiwa, Chiba 277-8582, Japan.}
\altaffiltext{2}{optical and Infrared Astronomy Division, National Astronomical Observatory, Mitaka, Tokyo
    181-8588, Japan.}
\altaffiltext{3}{Department of Astronomical Science, Graduate University for Advanced Studies (SOKENDAI), Mitaka,
    Tokyo 181-8588, Japan.}
\altaffiltext{4}{Observat\'{o}rio Nacional, Rua Jos\'{e} Cristino, 77. CEP 20921-400, S\~{a}o Crist\'{o}v\~{a}o,
    Rio de Janeiro-RJ, Brazil}
\altaffiltext{5}{Department of Physics, Graduate School of Science, The University of Tokyo, Bunkyo, Tokyo
    113-0033, Japan}
\altaffiltext{6}{Academia Sinica Institute of Astronomy and Astrophysics, P.O. Box 23-141, Taipei 10617, Taiwan}
\altaffiltext{7}{Research Center for Space and Cosmic Evolution, Ehime University, Matsuyama, Ehime 790-8577, Japan}
\altaffiltext{8}{Astronomical Institute, Tohoku University, Sendai, Miyagi 980-8578, Japan}
\altaffiltext{9}{Institute of Astronomy, National Tsing Hua University, No. 101, Section 2, Kuang-Fu Road, Hsinchu, Taiwan, 30013}
\altaffiltext{10}{Subaru Telescope, National Astronomical Observatory of Japan, 650 N Aohoku Pl, Hilo, HI 96720, USA}


\maketitle

\begin{abstract}
We conduct a systematic search for galaxy protoclusters at $z\sim3.8$ based on the latest internal data release
(S16A) of the Hyper SuprimeCam Subaru strategic program (HSC-SSP).
In the Wide layer of the HSC-SSP, we investigate the large-scale projected sky distribution of $g$-dropout
galaxies over an area of $121\,\mathrm{deg^2}$, and identify 216 large-scale overdense regions ($>4\sigma$
overdensity significance) that are good protocluster candidates.
Of these, 37 are located within $8\,\mathrm{arcmin}$ ($3.4\,\mathrm{physical\>Mpc}$) from other protocluster
candidates of higher overdensity, and are expected to merge into a single massive structure by $z=0$.
Therefore, we find 179 unique protocluster candidates in our survey.
A cosmological simulation that includes projection effects predicts that more than 76\% of these candidates will
evolve into galaxy clusters with halo masses of at least $10^{14}\,M_{\solar}$ by $z=0$.
The unprecedented size of our protocluster candidate catalog allowed us to perform, for the first time, an angular
clustering analysis of the systematic sample of protocluster candidates.
We find a correlation length of $35.0\,h^{-1}\,\mathrm{Mpc}$.
The relation between correlation length and number density of $z\sim3.8$ protocluster candidates is consistent
with the prediction of the $\Lambda$CDM model, and the correlation length is similar to that of rich clusters in
the local universe.
This result suggests that our protocluster candidates are tracing similar spatial structures as those expected
of the progenitors of rich clusters and enhances the confidence that our method to identify protoclusters at high
redshifts is robust.
In the coming years, our protocluster search will be extended to the entire HSC-SSP Wide sky coverage of
$\sim1400\,\mathrm{deg^2}$ to probe cluster formation over a wide redshift range of $z\sim2\mathrm{-}6$.
\end{abstract}

\section{INTRODUCTION}
Structure formation in the universe proceeds hierarchically.
Small perturbations in the initial density field grow and merge together over time, leading to the emergence of a
cosmic web: a filamentary structure with voids and overdensities \citep{bertschinger98}.
How galaxy clusters, the largest overdensities at the nodes of this web, are assembled remains an open question.
In which respect does the evolution of galaxies residing in these highest density peaks differ from that of those
in the field?
To address this question, it is necessary to directly investigate all the evolutionary stages of galaxy clusters
through cosmic time, including their progenitors, ``protoclusters" \citep{overzier16}.
In the local universe, galaxy clusters exhibit tight red sequences, composed of massive quiescent galaxies, and can
be traced by X-ray emission from the hot gas trapped in their massive dark matter halos.
Increasing redshifts, the fraction of star-forming galaxies becomes higher, and quiescent galaxies would be a rare
galaxy population even in overdense environments at $z\gtrsim2$.
Galaxies in protoclusters are predicted to be the first to transition from star-forming to quiescent due to
environmental effects.
Numerical simulations suggest that at early epochs these galaxies may experience enhanced accretion and galaxy
merger rates \citep{somerville15}.
Protoclusters thus represent a unique laboratory for the study of early (massive) galaxy growth.
However, direct empirical data in support of these expectations remains elusive.
Despite their importance, protoclusters remain poorly understood due to their rarity.
So far, the number of known protoclusters in the early universe ($z\gtrsim3$) is limited to less than a dozen
systems \citep[e.g.,][]{venemans07,toshikawa16}.

To find such rare protoclusters at high redshifts, radio galaxies (RGs) and quasars (QSOs) are commonly used as
tracers of protoclusters \citep[e.g.,][]{wylezalek13,adams15} because these galaxies hosting powerful active
galactic nuclei (AGN) are thought to be embedded in massive dark matter halos.
However, not all RGs and QSOs appear to reside in high-density environments
\citep{hatch11,mazzucchelli17,uchiyama17}, and \citet{hatch14} found that although RGs are found in environments
that are, on average, relatively biased compared to other galaxies, they can be found in a wide range of low- to
high-density environments based on a study of 419 RGs.
It is therefore still unclear if, and how the AGN activity in, for example, RGs and QSOs is related to their
surrounding environments.
Because of non-unity AGN duty cycles, we also expect that a large fraction of protoclusters would be missed if
using powerful AGNs as tracers.
Furthermore, strong radiation from RGs and QSOs could also result in a suppression of galaxy formation
\citep[e.g.,][]{barkana99,kashikawa07}.
Therefore, in order to avoid possible selection biases, it has been a longstanding goal to search for protoclusters
in large ``blank'' field.
At $z\sim1\mathrm{-}2$, many galaxy clusters have been discovered by blank surveys \citep[e.g.,][]{rettura14}.
Beyond $z\sim3$, while some protoclusters that do not host RGs or QSOs have been discovered
\citep{steidel98,ouchi05}, the number of such protoclusters is still very small due to very limited sky coverage
at a sufficient depth.
Based on a relatively wide $4\,\mathrm{deg^2}$ optical imaging survey of the Canada-France-Hawaii Telescope Legacy
Survey Deep Fields and follow-up spectroscopy, the number density of protoclusters was found to be only
$\sim1.5\,\mathrm{deg^{-2}}$ at $z\sim4$ \citep{toshikawa16}.

Here, we present a systematic survey of protoclusters at $z\sim3.8$ based on the wide-field survey with the Hyper
SuprimeCam \citep[HSC:][]{miyazaki12} conducted as part of the Subaru strategic program (SSP).
The HSC is mounted at the prime focus of the Subaru telescope and has a large field-of-view (FoV) of
$1.8\,\mathrm{deg^2}$.
The wide-field imaging capability of the HSC enables us to find high-redshift protoclusters in large blank fields
with relative ease.
The HSC-SSP started in early 2014 and will be completed by 2019.
The HSC-SSP is composed of three layers, the UltraDeep (UD; $3.5\,\mathrm{deg^2}$, $i\sim28\,\mathrm{mag}$),
the Deep ($26\,\mathrm{deg^2}$, $i\sim27\,\mathrm{mag}$), and the Wide layers ($1400\,\mathrm{deg^2}$,
$i\sim26\,\mathrm{mag}$) \citep{aihara17}.
By using the extremely wide-area coverage and the high sensitivity through five optical broad-bands ($g$-, $r$-,
$i$-, $z$-, and $y$-bands), we construct a systematically selected sample of protoclusters at $z\gtrsim3$.
The present paper is one in the series of papers on Lyman break galaxies (LBGs) based on the HSC-SSP, named Great
Optically Luminous Dropout Research Using Subaru HSC (GOLDRUSH), which include sample selection and UV luminosity
function \citep{ono17}, clustering analysis \citep{harikane17}, and protocluster search presented in this paper.
In addition, \citet{uchiyama17} and \citet{onoue17} discuss the relation between LBG overdensity and QSOs by using
the sample of protocluster candidates constructed in this paper.
The HSC-SSP also allows us to study high-redshift Ly$\alpha$ emitters (LAEs) by narrow-band imaging as the project
of Systematic Identification of LAEs for Visible Exploration and Reionization Research Using Subaru HSC
\citep[SILVERRUSH:][]{ouchi17,shibuya17a,shibuya17b,konno17}.
This paper is organized as follows.
Section \ref{sec:pcl} describes the selection of $z\sim3.8$ galaxies and the systematic sample of protocluster
candidates.
In Section \ref{sec:acf}, we investigate the spatial distribution of protocluster candidates through a clustering
analysis.
We will discuss the dark matter halo mass of protoclusters based on a systematic sample of protocluster candidates
in Section \ref{sec:dmhm}.
The summary is given in Section \ref{sec:sum}.
We assume the following cosmological parameters: $\Omega_\mathrm{M}=0.3$, $\Omega_\Lambda=0.7$,
$H_0=100h\mathrm{\,km\,s^{-1}\,Mpc^{-1}}=70\mathrm{\,km\,s^{-1}\,Mpc^{-1}}$, and magnitudes are given in the AB
system.
\begin{table*} [htb]
\caption{Fields in the HSC-SSP S16A data release. \label{tab:data}}
\begin{center}
\begin{tabular}{l|l|l|l} \hline
Name & R.A. & Decl. & effective area [deg$^2$] \\
\hline \hline
Wide-XMM & $1^{h}36^{m}00^{s}$ - $3^{h}00^{m}00^{s}$ & $-6^{\circ}00'00''$ - $-2^{\circ}00'00''$ & 31.3 \\
Wide-WIDE12H & $11^{h}40^{m}00^{s}$ - $12^{h}20^{m}00^{s}$ & $-2^{\circ}00'00''$ - $2^{\circ}00'00''$ & 17.0 \\
Wide-GAMA15H & $14^{h}00^{m}00^{s}$ - $15^{h}00^{m}00^{s}$ & $-2^{\circ}00'00''$ - $2^{\circ}00'00''$ & 39.3 \\
Wide-HECTOMAP & $15^{h}00^{m}00^{s}$ - $17^{h}00^{m}00^{s}$ & $42^{\circ}00'00''$ - $45^{\circ}00'00''$ & 12.6 \\
Wide-VVDS & $22^{h}00^{m}00^{s}$ - $23^{h}20^{m}00^{s}$ & $-2^{\circ}00'00''$ - $3^{\circ}00'00''$ & 20.7 \\
\hline
\end{tabular}
\end{center}
\end{table*}

\section{PROTOCLUSTER CANDIDATES \label{sec:pcl}}
\subsection{Selection of $z\sim3.8$ Galaxies}
In this paper, we make use of the latest internal HSC-SSP data release (S16A).
The current UD and Deep layers have not yet reached the final depths of the five-year HSC-SSP.
In the Wide layer, the survey area at full depth has been steadily increasing as the HSC-SSP proceeds.
The sky coverage with all five broad-bands ($g$, $r$, $i$, $z$, and $y$) is $178\,\mathrm{deg^2}$ in the Wide
layer of the S16A data release, and already dozens of times larger than any previous survey capable of detecting
protoclusters at $z\gtrsim3$.
Therefore, in this paper, we focus on a systematic search for protoclusters at $z\sim3.8$ in the Wide layer as our
initial attempt.
Although the Wide layer is composed of three large regions (two around the spring and autumn equator and the
other around the Hectmap region) at the end, the current Wide data release includes six disjoint regions in the
XMM-LSS, GAMA09H, WIDE12H, GAMA15H, HECTOMAP, and VVDS fields.

Image reduction, object detection, and photometry were performed by the reduction pipeline of the HSC-SSP (hscPipe;
see the details in \citet{bosch17}).
From this HSC-SSP catalog in the Wide layer, we select $z\sim3.8$ galaxy candidates using the Lyman break technique
($g$-dropout galaxies).
The construction of dropout galaxies is described in \citet{ono17}, and this study of protocluster search is
conducted based on this catalog though we apply some modifications on the sample selection in order to meet our
requirements.
We use the following color-selection criteria \citep{burg10,toshikawa16}, which are the same with those of
\citet{ono17}:
\[ 1.0<(g-r) \>\wedge\> -1.0<(r-i)<1.0 \>\wedge\> 1.5(r-i)<(g-r)-0.80. \]
In order to accurately estimate their color, we only use objects detected at more than $5\sigma$ and $3\sigma$
significance in the $i$- and $r$-bands, respectively; a $3\sigma$ limiting magnitude for the $g$-band was used in
the $g-r$ color if objects were fainter than the $3\sigma$ limiting magnitude in the $g$-band.
We use the {\sf CModel} magnitude, which is determined by fitting two-component, PSF-convolved galaxy models (de
Vaucouleurs and exponential), to estimate color.
For point sources, the CModel magnitude is consistent with a PSF magnitude.
Our color criteria are sensitive to galaxies in the redshift range of $z\sim3.3\mathrm{-}4.2$, corresponding to
$\Delta z \sim 740\,\mathrm{comoving\>Mpc}$.
The locus of contaminants (e.g., dwarf stars and passive galaxies at $z\sim0.5$) lies far from the selection
region on the two-color diagram composed of $g-r$ and $r-i$ \citep{toshikawa16}.
In addition to these color criteria, we remove objects which are located at the edge of images, cosmic rays,
saturated, and bad pixels by using the flags {\sf flags\_pixel\_edge}, {\sf flags\_pixel\_interpolated\_center},
{\sf flags\_pixel\_saturated\_center}, {\sf flags\_pixel\_cr\_center}, {\sf flags\_pixel\_bad}, which are products
of hscPipe indicating the reliability of the measurements.
Objects near bright stars are also masked by means of the flags {\sf flags\_pixel\_bright\_object\_center} and
{\sf flags\_pixel\_bright\_object\_any}.
It should be noted that the mask used around bright stars is very large ($\gtrsim1\,\mathrm{deg}$ diameter) for
very bright stars in the current version of the HSC-SSP catalog.
From these criteria of colors, detection significances, and measurement flags, nearly one million of $g$-dropout
galaxies are obtained in all six regions of the Wide layer, with number counts that are consistent with those of
\citet{burg10} (see \citet{ono17} for detail).

\subsection{Identification of Protocluster Candidates \label{sec:pclid}}
\begin{figure}
\includegraphics[width=8cm]{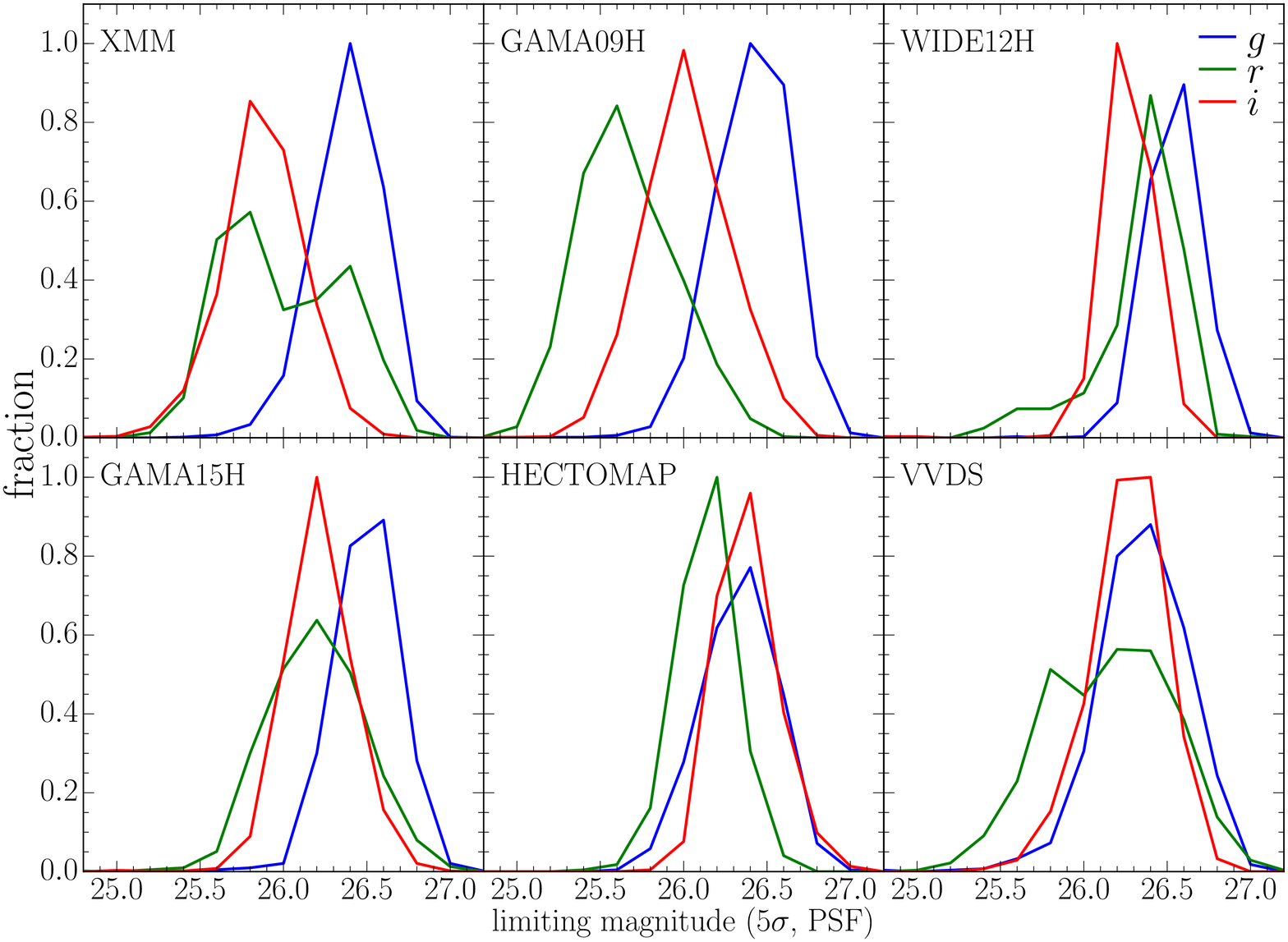} 
\caption{Sky noise distribution over the whole Wide layer.
    The blue, green and red lines indicate the sky noise measured in the $g$-, $r$-, and $i$-bands, respectively.}
\label{fig:limmag}
\end{figure}

The depth in the Wide layer is inhomogeneous over the whole area because the sky conditions vary during the
long-term observations; furthermore, even in the same portion of the Wide layer, all five optical filters are not
observed in the same night.
The inhomogeneity of depth could make a large impact on the number of detected objects, making it difficult to
fairly compare number density among different fields.
The hscPipe processes HSC imaging data separately in $1.7\times1.7\,\mathrm{deg^2}$ rectangular tracts, and a tract
is further divided into sub-regions of $12\times12\,\mathrm{arcmin^2}$ patches.
We estimate the sky noise of each patch, determined as the standard deviation of sky flux measured by the PSF
photometry (Figure \ref{fig:limmag}). 
We search for protocluster candidates in area where the $5\sigma$ limiting magnitudes of the $g$-, $r$-, and
$i$-band are fainter than 26.0, 25.5, and $25.5\,\mathrm{mag}$, respectively.
These limits, slightly shallower than the target depths, are met by most of the patches in the Wide layer except
the Wide-GAMA09H regions, whose $r$-band depth is shallower than the limit as shown in Figure \ref{fig:limmag}, and
the difference of depth between $r$- and $i$-bands is large.
Since the UV slope, or equivalently $r-i$ color, of $g$-dropout galaxies is expected to be flat, the imbalance
between $r$- and $i$-bands could bias the $g$-dropout selection.
Thus, the Wide-GAMA09H region is not used in this study.
The total effective area of our analysis is $121\,\mathrm{deg^2}$ (Table \ref{tab:data}).
In these regions, we use $g$-dropout galaxies down to the $i$-band magnitude of $25.0\,\mathrm{mag}$, corresponding
to the characteristic UV magnitude at $z\sim4$, for the estimate of surface number density.

\begin{figure*}
\includegraphics[width=16cm]{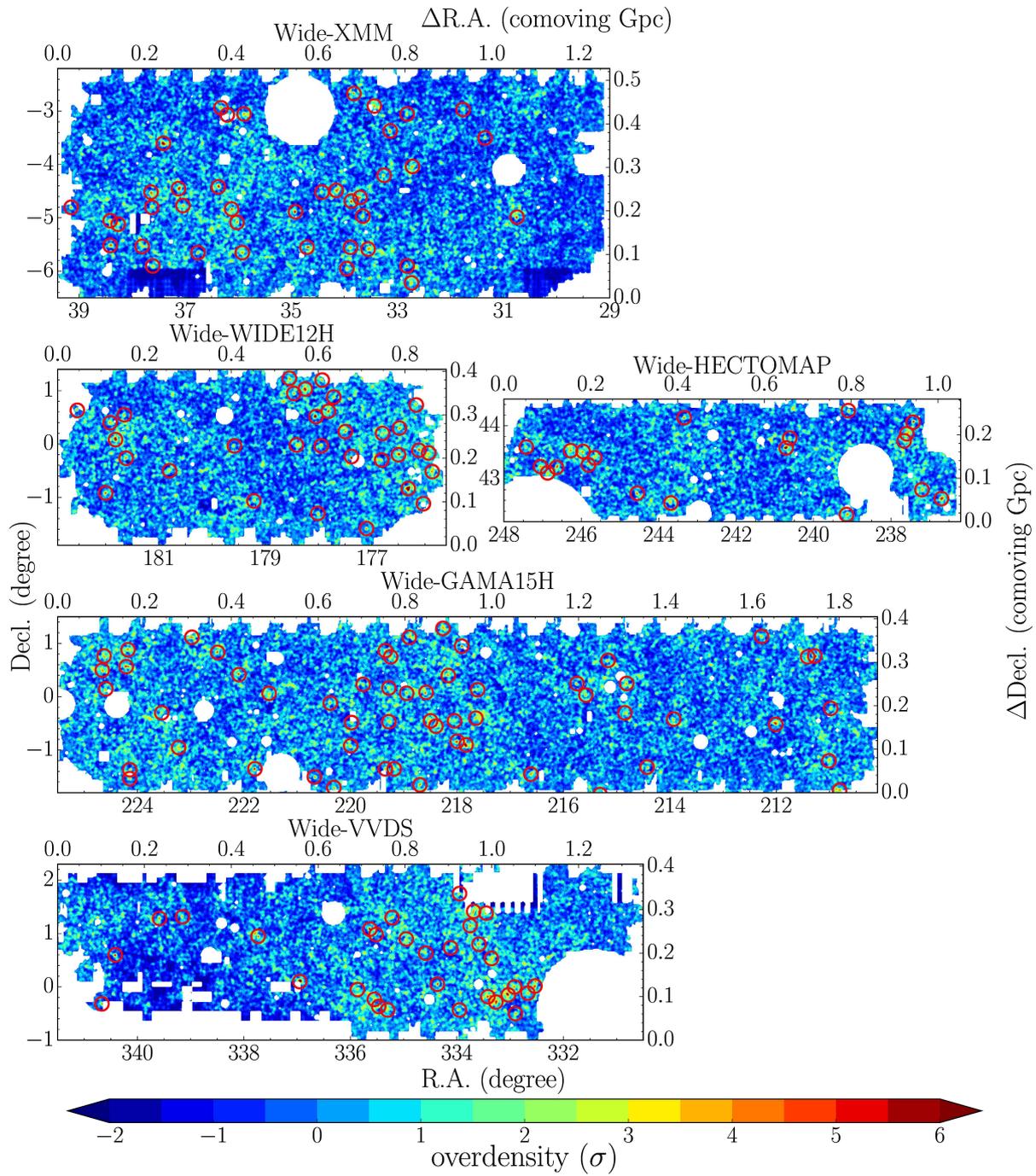} 
\caption{Overdensity contours of $g$-dropout galaxies.
    Higher density regions are indicated by redder colors. 
    The white regions are regions that are masked due to survey edges, bright stars, or insufficient depth.
    The positions of the unique 179 protocluster candidates are indicated by red circles.
    The area of all panels are shown at the same scale.}
\label{fig:cntr}
\end{figure*}

The local surface number density of $g$-dropout galaxies is quantified by following the same method as in
\citet{toshikawa16}, which is by counting the number of $g$-dropout galaxies within an aperture of
$1.8\,\mathrm{arcmin}$ ($0.75\,\mathrm{physical\>Mpc}$) radius.
This radius is expected to be the typical extent of regions which will coalesce into a single massive halo of
$>10^{14}\,M_{\solar}$ by $z=0$ \citep[e.g.,][]{chiang13}.
Although progenitors of more massive galaxy clusters could be even larger, they are still expected to exhibit
significant excess number density on a $\sim1.8\,\mathrm{arcmin}$ scale.
On the other hand, our choice may make it difficult to identify smaller structures, like progenitors of galaxy
groups, because their clustering signal would be diluted by applying somewhat larger apertures than their typical
extent.
Furthermore, projection effects resulting from the relatively wide redshift range of $g$-dropout galaxies will also
affect the estimate of surface number density because the typical size of protoclusters ($\Delta z\lesssim0.03$)
is tens times smaller than the redshift range of $g$-dropout galaxies.
It is possible that an alignment of two small structures by chance shows significantly high density, while the
overdensity of protoclusters can diminish due to fore/background void regions.
The apertures are distributed over the Wide layer in a grid pattern with intervals of $1\,\mathrm{arcmin}$.
By using only apertures in which the masked area is less than 5\%, the mean and the dispersion, $\sigma$, of the
number of $g$-dropout galaxies in an aperture are found to be 6.4 and 3.2, respectively.
The difference of the mean among the five fields of the Wide layer is only 0.54 ($0.17\sigma$) at most.
Thus, while the limiting magnitudes of each field are slightly different, it is confirmed that we obtain a highly
uniform sample over all five regions by limiting the selection to $g$-dropout galaxies to be brighter than
$25.0\,\mathrm{mag}$ and removing patches in which the $g$-, $r$-, and $i$-band depths are shallower than our
criteria. 
Both statistical errors and cosmic variance should contribute to the estimate of $\sigma$
\citep{somerville04}, but the derived $\sigma$ is almost equal to that expected from Poisson statistics
\citep{gehrels86} due to the small number of $g$-dropout galaxies in an aperture.
The detailed study of cosmic variance at high redshift will be addressed based on the HSC-Deep layer.
In this study, we only aim to identify significantly overdense regions rather than to accurately map environments
from low to high density.
The surface number density of $g$-dropout galaxies in masked regions is assumed to be the same as the mean, but
apertures in which $>50\%$ area is masked are excluded in our protocluster search.
Overdensity is defined as the excess surface number density from the mean, and overdensity contours of the Wide
layer are plotted in Figure \ref{fig:cntr}.

\begin{figure}
\includegraphics[width=8cm]{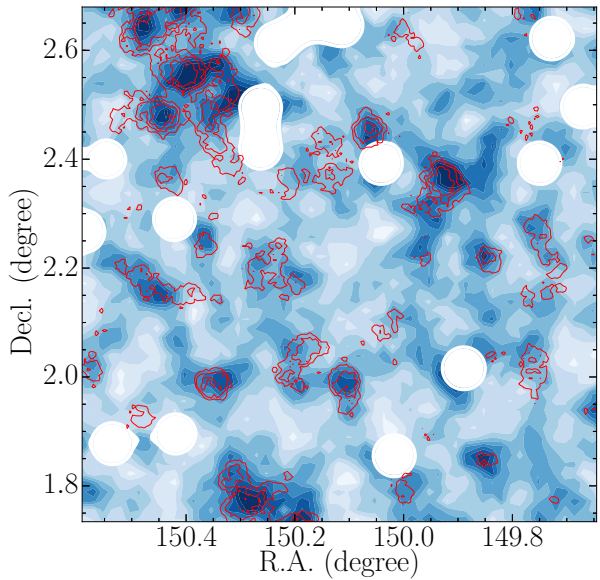} 
\caption{Overdensity contours of $g$-dropout galaxies based on the HSC (blue color scale) and CFHTLS (red lines)
    dataset.
    The white regions are masked areas due to bright stars.
    The positions of overdensity peaks in the HSC dataset are almost identical to those found in the CFHTLS
    dataset.}
\label{fig:sanity}
\end{figure}

Since the central $\sim1\,\mathrm{deg^2}$ area of the UD-COSMOS overlaps with the Canada-France-Hawaii Telescope
Legacy Survey (CFHTLS) Deep 2 field, we can make a consistency check of our overdensity estimate by directly
comparing with the overdensity contours of $g$-dropout galaxies in the CFHTLS \citep{toshikawa16}.
The same color selection of $g$-dropout galaxies and overdensity estimate described above are applied to both the
HSC and CFHTLS dataset.
As shown in Figure \ref{fig:sanity}, the overdensity contours derived by the HSC dataset are clearly consistent
with that of the CFHTLS dataset, suggesting that we can correctly map overdensity in the Wide layer as well.

\begin{figure}
\includegraphics[width=8cm]{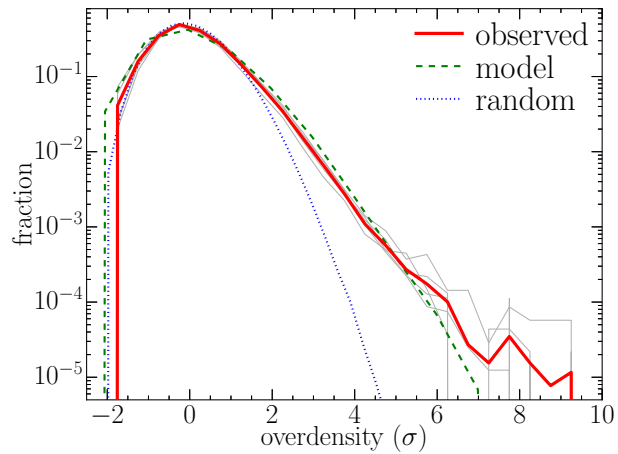} 
\caption{Distribution of overdensity significance.
    The gray and red lines indicate the individual regions of the Wide layer and their average, respectively.
    The green dashed and blue dotted lines show the distributions expected from our theoretical model and a Poisson
    distribution, respectively.}
\label{fig:dist}
\end{figure}

The distribution of overdensity significance in the Wide layer largely deviates from the expected distribution
from the Poisson distribution at the high-overdensity end (Figure \ref{fig:dist}).
For example, the number of $4\sigma$ overdense regions is $\sim30$ times higher than the expectation from Poisson
distribution.
Thus, high overdense regions would be resulted from clustering of physically associated galaxies rather than
chance alignment due to projection effect.
Furthermore, we have also compared with a theoretical prediction derived from the light-cone model of
\citet{henriques12} by selecting mock $g$-dropout galaxies and applying the same overdensity measurements (see
\S3.2 in \citet{toshikawa16} for details).
The observed distribution is consistent with that of the model prediction, although the model somewhat
underestimates the number of $>6\sigma$ overdensity significance regions.
Such overdense regions are extremely rare, and the excess observed in the Wide layer over the simulations comes as
no surprise given that the volume of our survey is $\sim1\,\mathrm{Gpc^3}$ compared to the volume of
$\sim0.4\,\mathrm{Gpc^3}$ for the simulation from which the light-cone models are extracted.
The model thus reasonably reproduces the observed overdensity distribution except at the extreme high-density end,
and the mass of descendant halo at $z=0$ can be deduced from the halo merger trees given by the simulation.
As shown in \citet{chiang13} and \citet{toshikawa16}, the surface overdensities observed toward protoclusters are
statistically correlated with the descendant halo mass at $z=0$ though we expect a large scatter caused by the
redshift uncertainty of the dropout galaxies (see \S3.2 in \citet{toshikawa16} for details).
Protocluster candidates are defined as regions where the overdensity significance is $>4\sigma$ at the peak.
With this definition, 76\% of these candidates are expected to evolve into galaxy clusters of 
$>10^{14}\,M_{\solar}$ at $z=0$, and the overdensities of the others are enhanced by projection effects though
they will be smaller structures than galaxy clusters.
On the contrary, only $\sim6\%$ of progenitors of $>10^{14}\,M_{\solar}$ halos at $z=0$ are expected to be located
on $>4\sigma$ regions at $z\sim4$ because the overdensity of most of progenitors are decreased by the projection
effects.
Our selection method for protocluster candidates can make a clean sample with high purity, though its completeness
is small.
It should be noted that our sample has low contamination, guaranteeing minimal effects of contamination on our
measurements of angular clustering (Section \ref{sec:acf}).
Because our selection strategy minimises the selection of structures suffering from large projection effects, our
sample of protocluster candidates will be biased to the richest structures with the average descendant halo mass
of $\sim5\times10^{14}\,M_{\solar}$.

\begin{figure}
\includegraphics[width=8.2cm]{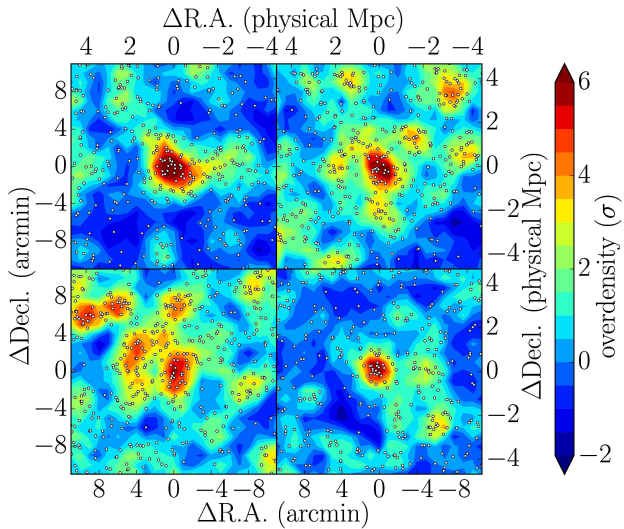} 
\caption{Examples of the protocluster candidates.
    The points indicate $g$-dropout galaxies.
    Individual protocluster candidates exhibit unique shapes, and some are accompanied by several other overdense
    regions (e.g., bottom-left and top-right panels).}
\label{fig:pcl}
\end{figure}

We have found 216 $>4\sigma$ overdense regions of $g$-dropout galaxies in the Wide layer.
The individual overdense regions show a wide range of morphologies (Figure \ref{fig:pcl}). 
The morphology of these overdense regions will eventually provide clues to understand galaxy/halo assembly from
the large-scale structure of the universe, but this is beyond the scope of this paper.
Some overdense regions have neighboring overdense regions within a few arcmin (Figure \ref{fig:pcl}).
Although, as mentioned above, $0.75\,\mathrm{physical\>Mpc}$ ($1.8\,\mathrm{arcmin}$) is the typical extent of
protoclusters, protocluster galaxies can be located a few or more physical Mpc away from their centers depending
on the direction of filamentary structure \citep[e.g.,][]{muldrew15}.
It is unlikely that two overdense regions are located within a few arcmin just by chance because the mean
separation is $\sim40\,\mathrm{arcmin}$ based on the surface number density of overdense regions.
\citet{toshikawa16} quantitatively investigated how far protocluster members are typically spread from the center
and found that galaxies lying within the volume of $R_\mathrm{sky}<8(6)\,\mathrm{arcmin}$ and $R_z<0.013(0.010)$
at $z\sim3.8$ will be members of the same protocluster with a probability of $>50(80)\%$.
Overdense regions which are located near each other are expected to merge into a single structure by $z=0$.
In this study, if $>4\sigma$ overdense regions are located within $8\,\mathrm{arcmin}$ from another more overdense
region, they can be regarded as the substructures of that protocluster though spectroscopic follow-up will be
required to distinguish from chance alignment.
Of 216 $>4\sigma$ overdense regions, 37 have neighboring more overdense regions.
The fraction of neighboring overdense regions is significantly higher than that expected by uniform random
distribution ($N=10.6\pm3.2$), implying that the large fraction of neighboring overdense regions are physically
associated with each other rather than chance alignment.
As a result, we have found 216 protocluster candidates at $z\sim3.8$, and 179 of them would trace the unique
progenitors of galaxy clusters in the Wide layer, which is about ten times larger than any previous study of
protoclusters ($N\sim10\mathrm{-}20$ at $z\gtrsim3$).
According to \citet{toshikawa16}, three out of four protocluster candidates identified by the same method are
confirmed to be real protoclusters by spectroscopic follow-up observations, which is consistent with the model
prediction.

\section{ANGULAR CLUSTERING \label{sec:acf}}
\begin{figure}
\includegraphics[width=8cm]{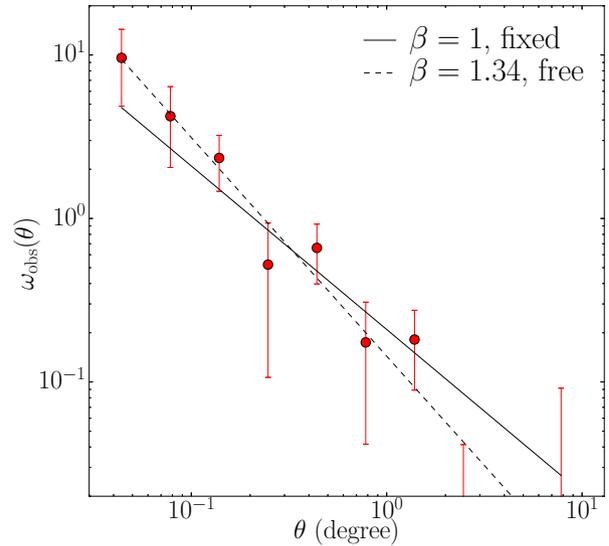} 
\caption{Observed angular correlation function of all protocluster candidates.
    The solid and dashed lines show the best-fitted power laws in the case of fixed and free $\beta$, respectively.}
\label{fig:acf}
\end{figure}

Based on the systematic sample produced by the HSC-SSP, we investigate the spatial distribution of protocluster
candidates at $z\sim3.8$ through the angular correlation function, $\omega(\theta)$.
In order to include any small-scale structure in the correlation function, in this analysis we use all $>4\sigma$
overdense regions instead of only unique protocluster candidates.
We measure the observed $\omega(\theta)$ using the estimator presented in \citet{landy93}:
\begin{equation}
\omega_\mathrm{obs}(\theta) = \frac{DD(\theta) -2DR(\theta) +RR(\theta)}{RR(\theta)},
\end{equation}
where DD, DR, and RR are the number of unique data-data, data-random, and random-random pairs with angular
separation between $\theta-\Delta\theta/2$ and $\theta+\Delta\theta/2$, respectively.
As shown in Figure \ref{fig:pcl}, the overdense regions are generally found to have $3\mathrm{-}6\,\mathrm{arcmin}$
extents within $\gtrsim2\sigma$ regions and show various, complex shapes.
The coordinate of overdense regions is simply defined as the position of their overdensity peak.
The locations of surface overdensity peaks can be affected by projection effects, but the typical uncertainty is
expected to be only $0.5\,\mathrm{arcmin}$ ($\sim2\,\mathrm{arcmin}$ at worst) by using theoretical models
\citep{toshikawa16}.
We distribute 40,000 random points in the same geometry as protocluster candidates.
The uncertainty of $\omega_\mathrm{obs}(\theta)$ is estimated using the bootstrap method as follows.
We randomly select our protocluster candidates, allowing for redundancy, and calculate
$\omega_\mathrm{obs}(\theta)$.
This calculation is repeated 100 times, and the uncertainty of each angular bins is determined by the root mean
square of all of the bootstrap steps.
Figure \ref{fig:acf} shows the angular correlation function for all $>4\sigma$ overdense regions at $z\sim3.8$ in
the Wide layer.

The angular correlation function can be parameterized by a power law: $\omega(\theta) = A_\omega\theta^{-\beta}$.
The slope, $\beta$, is found to be $\sim1.0$, which does not strongly depend on redshift and mass of clusters at
$z\lesssim2$ \citep[e.g.,][]{bahcall03,papovich08}.
We use a least-square technique to fit a power-law function to the angular correlation function.
The observed angular correlation function is biased to lower amplitude because the size of the survey field is
finite.
Thus, the true angular correlation function is estimated by adding a constant value, known as the integral
constraint:
\begin{equation}
\omega(\theta) = \omega_\mathrm{obs}(\theta) + \frac{\Sigma A_\omega\theta^{-\beta}RR(\theta)}{\Sigma RR(\theta)}
    + \frac{1}{N},
\end{equation}
where $N$ is the number of objects.
Since the angular correlation function, $\omega(\theta)$, is the projected three-dimensional spatial correlation
function, $\xi(\theta)$, we can derive $\xi(\theta)$ from $\omega(\theta)$ if the redshift distribution of
protocluster candidates is known \citep{limber53,phillipps78}.
The spatial correlation function can also be parameterized by a power law of $\xi(r)=(r/r_0)^{-\gamma}$.
The slope, $\gamma$, is related to $\beta$ as $\gamma=\beta+1$, and the correlation length, $r_0$, is estimated
from the following equation:
\begin{equation}
r_0 = \Biggl[ \frac{A_\omega}{H_\gamma} \frac{\bigl(\int\frac{dN}{dz}dz\bigr)^2}
    {\int f(z) d_C^{1-\gamma}(z)\bigl(\frac{dN}{dz}\bigr)^2 E(z)dz}\Biggr]^{1/\gamma}, \label{eq:limber}
\end{equation}
where $f(z)$ describes the redshift dependence of $\xi(r)$, $d_C$ is the comoving distance, $dN/dz$ is the redshift
selection function, $E(z)$ and $H_\gamma$ are defined as:
\begin{eqnarray}
E(z) &=& \frac{H_0}{c} \biggl[ \Omega_m(1+z)^3+\Omega_\Lambda\biggr]^{1/2} \\
H_\gamma &=& \sqrt{\pi}\frac{\Gamma[(\gamma-1)/2]}{\Gamma(\gamma/2)}.
\end{eqnarray}
We use $f(z)=[(1+z)/(1+z_c)]^{-(3+\epsilon)}$, where $z_c$ is the typical redshift of protocluster candidates and
$\epsilon=1.2$ \citep{roche99}.
Solving Equation \ref{eq:limber}, we assume that the redshift selection function of protocluster candidates
corresponds to that of $g$-dropout galaxies.
The redshift selection function is derived by the same method as in \citet{toshikawa16}.
In the case $\beta$ is fixed to 1.0, we derive $r_0=35.0^{+3.0}_{-3.3}\,h^{-1}\,\mathrm{Mpc}$.
When both $\beta$ and $r_0$ are free parameters, the best-fitted values are found to be $\beta=1.34\pm0.20$ and
$r_0=35.7^{+4.6}_{-5.5}\,h^{-1}\,\mathrm{Mpc}$.
Although there is a slight difference in the value of $\beta$, the $r_0$ derived in the case of fixed $\beta$
corresponds to that in the case of free $\beta$ within $1\sigma$ uncertainty.
It should be noted that we assume that the redshift distribution of protoclusters is identical to that of
$g$-dropout galaxies, though it is unlikely that protoclusters are found at the lower- or higher-ends of the
redshift selection function of $g$-dropout galaxies; thus, the redshift distribution of protoclusters may be
narrower.
If there are systematical differences of physical properties between protocluster and field galaxies, their
redshift tracks on two-color diagram and redshift distribution could also be different.
So far, significant differences between protocluster and field galaxies have not been found at $z\gtrsim4$ though
only a few protoclusters have been investigated to date \citep{overzier09}.
To evaluate the redshift distribution of protoclusters is beyond this study, and further detailed studies are
required (e.g., systematic follow-up spectroscopy, or model comparison).

\section{HALO MASS ESTIMATE \label{sec:dmhm}}
The mass of dark matter halo is one of key quantities to characterize the physical properties of protoclusters
because the evolution of dark matter halo or the formation of the large-scale structure are relatively
well-understood compared with complex baryon physics.
Once the dark matter halo mass is estimated at high redshifts, we can predict the descendant halo mass in the
context of hierarchical structure formation model.
However, in actual observations, protocluster candidates are defined by the number density of galaxies, and it is
not straightforward to estimate their halo mass.
For local clusters, velocity dispersion is frequently used to measure halo mass, and the same technique applies to
protoclusters at high redshifts; however, protoclusters would be far from virialization.
Another method is based on galaxy number density, which can be converted into mass density by using the bias
parameter; then, mass can be calculated by the volume of a protocluster \citep[e.g.,][]{steidel98,venemans07}.
It should be noted that the mass estimated by this way means total mass which will collapse into a single structure
rather than current halo mass at high redshift.

In this study, we show two alternative approaches.
The one is abundance matching by assuming protoclusters occupy all most massive dark matter halos, and the other
is to utilize the clustering strength of protocluster candidates estimated in Section \ref{sec:acf}.
We use these method on protocluster candidates at $z\sim3.8$ for the first time because a systematic sample is
required.

\subsection{Abundance Matching}
\begin{figure}
\includegraphics[width=8cm]{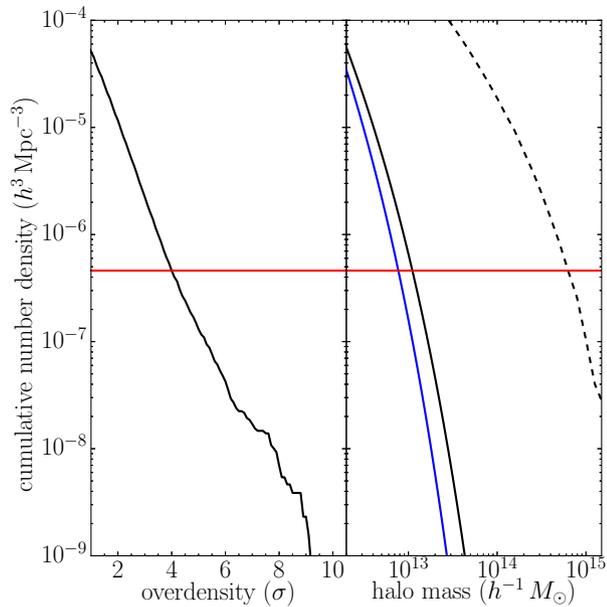} 
\caption{Cumulative number density of overdense regions of $g$-dropout galaxies (left) and dark matter halos
    (right).
    In the right panel, the solid lines represent halo mass functions at $z\sim3.8$ with different cosmology 
    (black: WMAP9, blue: Planck), and halo mass function at $z=0$ is shown by the dashed line.
    The horizontal line (red) indicates the number density of protocluster candidates at $z\sim3.8$ ($>4\sigma$
    overdense regions).
    It should be noted that the number density of descendant clusters at $z=0$ would be smaller than that of
    protoclusters at $z\sim3.8$ due to merging.}
\label{fig:mf}
\end{figure}

The number density of the protocluster candidates is estimated to be $n=4.6\times10^{-7}\,h^3\,\mathrm{Mpc^{-3}}$
by assuming that the redshift selection function of protoclusters is identical to that of $g$-dropout galaxies.
The number density of protocluster candidates is found to correspond to that of dark matter halos of
$\sim0.8\mathrm{-}1.1\times10^{13}\,h^{-1}\mathrm{M_\odot}$ at $z\sim3.8$ (Figure \ref{fig:mf}).
The difference of cosmology or mass function model does not have a large effect on the estimate of halo mass.
In the same manner, it would be possible to predict the descendant halo mass at $z=0$, though we need to consider
the change of the number density from $z\sim3.8$.
If individual protocluster candidates will evolve different clusters, or the number density at $z=0$ is identical
to that at $z\sim3.8$, the descendant halo mass of protocluster candidates is expected to be
$\sim6.3\times10^{14}\,h^{-1}\mathrm{M_\odot}$.
However, as discussed in Section \ref{sec:pclid}, some protocluster candidates have other neighboring protocluster
candidates.
Since such protocluster candidates can merge into single massive structures, the number density at $z=0$ will be
smaller than at $z\sim3.8$.
In case that protocluster candidates located within the half of mean separation ($20\,\mathrm{arcmin}$) will
coalesce into a single halo by $z=0$, the number density is decreased by 37\%.
Along with the decline of number density, the expected descendant halo mass is increased to
$\sim7.6\times10^{14}\,h^{-1}\mathrm{M_\odot}$.
As shown in Figure \ref{fig:mf}, the slope of mass function is very steep at the high-mass end; hence, the estimate
of halo mass based on abundance matching have a small dependence on the fraction of protocluster merging.

\subsection{Clustering}
We also calculated the dark matter halo mass of our protocluster candidates based on our clustering analysis.
We have used the analytical model proposed by \citet{sheth01} and \citet{mo02}, which has been shown to provide a
connection between the bias of the dark matter halo and the mass of the dark matter halo based upon ellipsoidal
collapse \citep{sheth01}, provided the bias of the dark matter halo and the redshift are given.
We estimate the mean dark matter halo mass, $\langle M_h \rangle $, from a comparison between the effective bias
parameter and the bias of the dark matter halo.
The dark matter halo mass was estimated as
$\langle M_h \rangle =2.3^{+0.5}_{-0.5}\times 10^{13}\,h^{-1}\mathrm{M_\odot}$, which is roughly consistent with
that estimated by the abundance matching.
According to this dark matter halo mass at $z\sim3.8$, the descendant halo mass at $z=0$ is evaluated as
$\langle M_h \rangle =4.1^{+0.7}_{-0.7}\times 10^{14}\,h^{-1}\mathrm{M_\odot}$ using the extended Press-Schechter
model.

\subsection{Relation between Abundance and Clustering Strength}
\begin{figure}
\includegraphics[width=8cm]{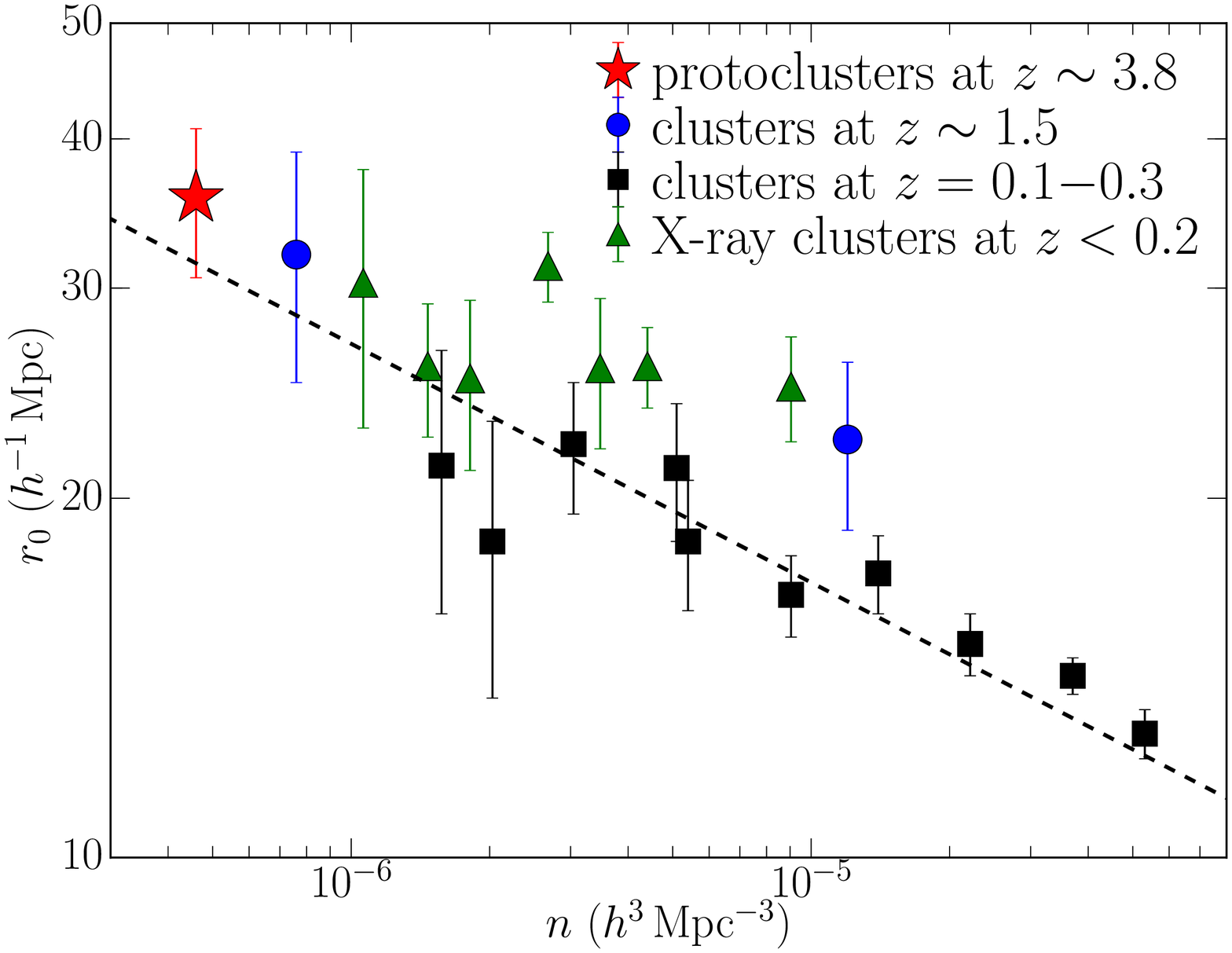} 
\caption{Comoving number density, $n$, versus correlation length, $r_0$, for protocluster candidates at $z\sim3.8$
    (red star, this study), distant clusters at $z\sim1.5$ \citep[blue circles,][]{papovich08,rettura14}, local
    clusters at $z=0.1\mathrm{-}0.3$ \citep[black squares,][]{croft97,bahcall03}, and X-ray-selected clusters at
    $z<0.2$ \citep[green triangles,][]{abadi98,collins00,bahcall03}.
    The dashed line indicates the expected relation between $n$ and $r_0$ from $\Lambda$CDM
    \citep[Equation \ref{eq:r0}:][]{younger05}.}
\label{fig:n-r0}
\end{figure}

Next, we investigate the relation between the spatial number density, $n$, and the correlation length of
protocluster candidates. 
\citet{younger05} found that the $n$-$r_0$ relation does not vary with redshift based on $N$-body dark matter
simulations.
The consistency in the $n$-$r_0$ relation across redshift can be explained as resulting from the fact that galaxy
clusters occupy the high-mass end of the halo mass function.
The formation of such massive structure started in the early universe, and the abundance of galaxy clusters or
their progenitors at high redshift does not dramatically vary with redshifts.
The correlation length of galaxy clusters is furthermore relatively stable assuming that they do not move very far
away from their initial location.
\citet{younger05} proposed an analytic approximation to the $\Lambda$CDM $n$-$r_0$ relation of
\begin{equation}
r_0 = 1.7 \biggl(\frac{n}{h^3\,\mathrm{Mpc^{-3}}}\biggr)^{-0.2}\,h^{-1}\,\mathrm{Mpc}. \label{eq:nr}
\label{eq:r0}
\end{equation}
According to Equation \ref{eq:nr}, the correlation length is calculated to be $r_0=31.4\,h^{-1}\,\mathrm{Mpc}$ from
the number density, which is consistent with the one derived from the clustering analysis (Figure \ref{fig:n-r0}).
The values of $r_0$ and $n$ found for our sample of protocluster candidates are comparable to those of
X-ray-selected galaxy clusters \citep[e.g.,][]{bahcall03}.
The correlation length is larger than that of optically-selected local clusters \citep{postman92} and Spitzer
selected clusters at $z\sim1$ \citep{papovich08}, but very similar to that of mid-infrared selected clusters at
$z\sim1.5$ in the South Pole Telescope Deep Field survey \citep{rettura14}.
This result suggests that our protocluster candidates are tracing very similar spatial structures as those expected
of the progenitors of rich clusters and enhances the confidence that our method to identify protoclusters at high
redshifts is robust.

\section{SUMMARY \label{sec:sum}}
We have presented a systematic sample of protoclusters at $z\sim3.8$ in the HSC-Wide layer based on the latest
internal data release of the HSC-SSP (S16A).
The unprecedentedly wide survey of the HSC-SSP allowed us to perform a large protocluster search without having to
rely on preselection of common protocluster probes (e.g., RGs or QSOs).
We selected a total of 216 overdense regions with an overdensity significance greater than $4\sigma$.
Of these, 37 can be considered to be substructures of a larger overdensity region, and thus in total we have
identified 179 unique protocluster candidates.
By comparing with theoretical models, we found that $>76\%$ of them are expected to evolve into galaxy clusters of
$>10^{14}\,M_{\solar}$ by $z=0$.
We investigated for the first time the spatial distribution of protoclusters through clustering analysis, and the
correlation length is found to be $r_0=34.1\,h^{-1}\,\mathrm{Mpc}$. 
Both the abundance matching method and the clustering analysis resulted in consistent protocluster halo masses of
$\sim1\mathrm{-}2\times10^{13}\,M_{\solar}$ at $z\sim3.8$.
These halos are expected to evolve into galaxy clusters of $\gtrsim5\times10^{14}\,\mathrm{M_\odot}$ by $z=0$
(Figures \ref{fig:mf} and \ref{fig:n-r0}).
The relation between number density and correlation length is consistent with the prediction of the $\Lambda$CDM
model, suggesting that our sample of protocluster candidates at $z\sim3.8$ indeed probes the progenitors of local
rich clusters.
When the HSC-SSP is completed, we expect that $>1000$ protoclusters will be identified at $z\sim3.8$ from the Wide
layer and a total of $\sim100$ protoclusters from $z\sim2$ to $z\sim6$ will be found from the UD and Deep layers.
Based on the systematic sample of protoclusters across cosmic time, we will be able to understand the process of
cluster formation from their birth to maturity.

\bigskip
\begin{ack}
The HSC collaboration includes the astronomical communities of Japan and Taiwan, and Princeton University.
The HSC instrumentation and software were developed by the National Astronomical Observatory of Japan (NAOJ), the
Kavli Institute for the Physics and Mathematics of the Universe (Kavli IPMU), the University of Tokyo, the High
Energy Accelerator Research Organization (KEK), the Academia Sinica Institute for Astronomy and Astrophysics in
Taiwan (ASIAA), and Princeton University.
Funding was contributed by the FIRST program from Japanese Cabinet Office, the Ministry of Education, Culture,
Sports, Science and Technology (MEXT), the Japan Society for the Promotion of Science (JSPS), Japan Science and
Technology Agency (JST), the Toray Science Foundation, NAOJ, Kavli IPMU, KEK, ASIAA, and Princeton University.
This paper makes use of software developed for the Large Synoptic Survey Telescope (LSST).
We thank the LSST Project for making their code available as free software at http://dm.lsst.org.
We thank the anonymous referee for valuable comments and suggestions that improved the manuscript.
We acknowledge supports from the JSPS grant 15H03645.
Authors RO and YTL received support from CNPq (400738/2014-7).

\end{ack}

\end{document}